\documentclass[aps,twocolumn,showpacs,superscriptaddress,citeautoscript,reprint,pre,floatfix]{revtex4-1}

\usepackage{bm}
\usepackage{graphicx}
\usepackage{amsmath}
\usepackage{amsfonts}
\usepackage{amssymb}
\usepackage{units}

\newcommand{\vc}[1]{\bm{#1}}

\begin{document}

\title[Chiral solitons in a non-linear model of helical molecules]%
{Chiral solitons in a non-linear model of helical molecules}

\author{P.~Albares}

\affiliation{NANOLAB, Dpto.\ de F\'{\i}sica Fundamental, Universidad de Salamanca, E-37008 Salamanca, Spain}

\author{E.~D\'{\i}az}

\affiliation{GISC, Departamento de F\'{\i}sica de Materiales, Universidad Complutense, E-28040 Madrid, Spain}

\author{Jose M.~Cerver\'o}

\affiliation{NANOLAB, Dpto.\ de F\'{\i}sica Fundamental, Universidad de Salamanca, E-37008 Salamanca, Spain}

\author{F.~Dom\'{\i}nguez-Adame}

\affiliation{GISC, Departamento de F\'{\i}sica de Materiales, Universidad Complutense, E-28040 Madrid, Spain}

\author{E.~Diez }

\affiliation{NANOLAB, Dpto.\ de F\'{\i}sica Fundamental, Universidad de Salamanca, E-37008 Salamanca, Spain}

\author{P. G.~Est\'{e}vez}

\affiliation{NANOLAB, Dpto.\ de F\'{\i}sica Fundamental, Universidad de Salamanca, E-37008 Salamanca, Spain}


\begin{abstract}

In this paper an effective integrable non-linear model describing the electron spin dynamics in a deformable helical molecule with weak spin-orbit coupling is presented. Non-linearity arises from the electron-lattice interaction and it enables the formation of a variety of stable solitons such as bright solitons, breathers and rogue waves, all of them presenting well defined spin projection onto the molecule axis. A thorough study of the soliton solutions is presented and discussed.

\end{abstract}

\maketitle

\section{Introduction} \label{sec:intro}

Solitons in organic helical molecules acquired a remarkably relevance after they were put forward as a possible explanation of how the energy released from the adenosine triphosphate protein is transfered across $\alpha$-helical proteins~\cite{Davydov79}. Today, they play a fundamental role to describe DNA denaturalization by way of the so called Peyrard-Bishop model~\cite{Peyrard89}. These models were also important for a complete analysis of the charge and energy transport properties of DNA molecules~\cite{Komineas02,Maniadis05,Diaz08}. In all these scenarios, it is assumed that organic molecules are quite deformable and consequently the interaction between quasiparticles (electrons or excitons) and the lattice vibrations is not negligible. On the contrary, such interaction enables the existence of stable solitons that propagate coherently along the molecule.

In the last decade, a large variety of experiments have shown the existence of chiral spin-selectivity in organic helical molecules~\cite{Goehler11,Xie11,Mishra11,Dor13,Kettner15,Mondal15,Einati15,Kiran16,Aragones17}. This effect results from the spin-orbit coupling (SOC) between the electronic momentum and the molecular electric field created by the helical arrangement of molecular dipoles. Many theoretical models have been proposed to explain these experimental evidences within different frameworks~\cite{Yeganeh09,Medina12,Gutierrez12,Guo12,Guo14a,Eremenko13,Rai13,Gutierrez13,Medina15,Caetano16}. However none of them was able to provide a good quantitative agreement with experimental data yet. Most recently, a few studies highlight the influence of the electron-lattice interaction on spin transport in organic helical molecules~\cite{Behnia16,Varela17,Diaz17bis}. This opens a new field of study within the scope of non-linear quantum dynamics that we further explore in this work. Our results will be also relevant for other physical systems, mainly Bose-Einstein condensates (BECs). Experiments performed in BECs have shown the versatility to study the influence of a helical SOC in non-linear systems~\cite{Lin11}. In this regard, there are several theoretical studies that predict the existence of a large variety of propagating solitons depending on the interacting parameters in BECs~\cite{Achilleos13,Wen16,Kartashov17,Sakaguchi17}.

In the following, we briefly introduce the non-linear model that describes the interaction of an electron moving along the axis of a deformable helical molecule. In addition to the kinetic energy, the electron undergoes an unconventional Rashba-like SOC arising from its motion in the helical arrangement of peptide dipoles~\cite{Diaz17,Diaz17bis}. The electron-lattice interaction adds an additional non-linear term to the Schr\'{o}dinger equation. We start by demonstrating the integrability of the model using the Painlev\'{e} test~\cite{Weiss}. We then turn to the main goal of the work, namely the detailed analysis of the solitons supported by the equation. For defocusing non-linear interaction, we find dark solitons that generalizes the solitons of the Manakov system~\cite{Manakov}. For focusing non-linear interactions, breathers and rogue waves are explicitly described. Furthermore, in this case solutions in terms of cnoidal waves also exist. The hyperbolic limit yields a generalization of the Davydov soliton~\cite{Davydov79}.

\section{Non-linear Schr\"{o}dinger equation for a deformable helical molecule} \label{sec:deformable}

Our starting point to describe the spin dynamics in helical molecules is the one-dimensional model introduced in Refs.~\cite{Diaz17,Diaz17bis}. In this model, the spin selectivity arises from an unconventional Rashba-like SOC, reflecting the helical symmetry of molecules due to the electron motion in a helical arrangement of peptides dipoles. To be specific, a helical conformation of tangentially oriented dipoles is considered to be spin-orbit coupled to the electron motion directed along the helical axis. 

Assuming that the helical molecule is oriented along the $Z$ axis, the resulting dimensionless Hamiltonian $H$ reads
\begin{equation}
H =- \begin{pmatrix} 
        \partial_{\xi\xi} & 2\pi\gamma\,e^{-i2\pi \xi}(i\partial_\xi+\pi)\\
        2\pi\gamma\,e^{i2\pi \xi}(i\partial_\xi-\pi) & \partial_{\xi\xi}
     \end{pmatrix}\ .
\label{eq:1}
\end{equation}
where energy is measured in units of $\hbar^2/2mb^2$, $m$ and $b$ being the electron mass and pitch of the helix, respectively. Here $\xi=z/b$, $\gamma$ stands for a dimensionless constant that is proportional to the magnitude of the SOC, and the subscript indicates differentiation with respect to $\xi$. In order to describe the electron-lattice interaction, we add a non-linear term to the Hamiltonian~(\ref{eq:1}). Such a  term is expected to appear within the adiabatic approximation, according to Davydov's theory~\cite{Davydov79}. The non-linear Schr\"{o}dinger equation (NLS) describing the dynamics of the spinor state $\vc{\chi}(\xi,t) = \left[\chi_\uparrow(\xi,t),\chi_\downarrow(\xi,t)\right]^{\intercal}$ reads
\begin{equation}
i \partial_t \vc{\chi}(\xi,t)=H\vc{\chi}(\xi,t)
+2g\left[\vc{\chi}^{\dag}(\xi,t)\cdot \vc{\chi}(\xi,t)\right]\vc{\chi}(\xi,t)\ ,
\label{eq:2}
\end{equation}
where $H$ is given in Eq.~(\ref{eq:1}). Hereafter we consider both $g>0$ (defocusing case) and $g<0$ (focusing one).

The integrability of this equation can be analyzed by using the Painlev\'{e} test~\cite{Weiss,Estevez07}. This test proves the integrability of Eq.~(\ref{eq:2}) and yields its three component Lax pair. The Painlev\'{e} property can be also used  to derive Darboux transformations and an iterative procedure for obtaining solutions. It can be also proved that Eq.~(\ref{eq:2}) is the only integrable case of a model very recently put forward by Kartashov and Konotov to study the dynamics of BECs with helical SOC~\cite{Kartashov17}. Furthermore, Eq.~(\ref{eq:2}) can be considered as a generalization of the Manakov system~\cite{Manakov,Vishnu}, which is often also called vector NLS system \cite{ablowitz}. Integrability properties of this Manakov system and the Painlev\'e property are describes in references \cite{Xing,Deng}. Different generalizations of this Manakov system can be found in \cite{latchio} and, more recently in \cite{zezyulin}.

\section{The singular manifold method} \label{sec:smm}

We start by rewriting Eq.~(\ref{eq:2}) in autonomous form through the changes
\begin{equation}
 \vc{\chi}=N_g 
      \begin{pmatrix}
         e^{-i\pi(\xi+\pi t)} & 0\\
          0 & e^{i\pi(\xi-\pi t)}  
      \end{pmatrix} \vc{\alpha}\ ,
  \label{eq:3}
\end{equation}
where $N_g=\sqrt{1/g}$ for $g>0$ (defocusing non-linear interaction) and $N_g=i\sqrt{1/\mid g\mid}$ for $g<0$ (focusing non-linear interaction). In both cases, the change yields the equations
\begin{eqnarray}
\left(i\partial_t+\partial_{\xi\xi}-2i\pi\partial_{\xi}
    -2\vc{\alpha}^{\dag}\!\!\cdot \vc{\alpha}\right)\alpha_1+ 2i\pi\gamma\partial_{\xi}\alpha_2&=0\ ,
    \nonumber\\
     2i\pi\gamma\partial_{\xi}\alpha_1+  \left(i\partial_t+\partial_{\xi\xi}+2i\pi\partial_{\xi}
    -2\vc{\alpha}^{\dag}\!\!\cdot \vc{\alpha}\right)\alpha_2&=0\ ,
    \label{eq:5}
\end{eqnarray}
where $\alpha_j$ with $j=1,2$ denotes the components of the spinor $\vc{\alpha}(\xi,t)$. In Ref.~\cite{Kartashov17}, Kartashov and Konotop proposed a one-dimensional non-linear model for moving solitons in a spatially inhomogeneous BEC with helical SOC. Is is not difficult to prove that the Gross-Pitaevskii equation for that model when the Zeeman splitting is negligible reduces to Eq.~(\ref{eq:5}).

A powerful tool for the study of the integrability of a system as Eq.~(\ref{eq:5}) is the Painlev\'e test  \cite{Weiss}, which requires to do the following ansatz for the components of $\vc \alpha$
\begin{equation}
\alpha_1=\sum_{j=0}^{j=\infty}a_j\phi^{j-1}\ , 
\quad\quad
\alpha_2=\sum_{j=0}^{j=\infty}b_j\phi^{j-1}\ .
\label{eq:7}
\end{equation}
This ansatz means that the solutions are single valued in the singularity manifold $\phi(\xi,t)=0$. The leading order analysis trivially yields
\begin{equation}
a_0=A\phi_{\xi}\ ,\quad\quad
b_0=B\phi_{\xi}\ ,
\label{eq:8}
\end{equation}
where $AA^{\dag}+BB^{\dag}=1$.

A straightforward calculation provides triple resonances in $j=0$ and $j=3$ and a single resonance in $j=4$. The symbolic calculations have been handled with MAPLE. The conditions at the resonances are identically satisfied. Therefore, we can conclude that the solutions are single valued around the singularity manifold and the equation has the Painlev\'{e} property. This Painlev\'{e} property is usually considered as a proof of the integrability of the equation, especially when it can be used to derive the linear spectral problem (Lax pair) associated to the equation~\cite{Estevez07}. The equivalence between the Painlev\'{e} property and the Lax pair can be achieved through the so called Singular Manifold Method (SMM)~\cite{Weiss,Estevez16}. This is the tool that we shall use in the rest of the paper to derive the main properties and solutions of the integrable non-linear system~(\ref{eq:5}). It is worth mentioning that the same Painlev\'{e} test, when applied to the model introduced by Kartashov and Konotop in Ref.~\cite{Kartashov17}, is only satisfied when the Zeeman splitting vanishes. Therefore, we are led to the conclusion that Eq~(\ref{eq:5}) is the only integrable case of the model given in Ref.~\cite{Kartashov17}.

The SMM implies the truncation of the series~(\ref{eq:7}) to the constant level
\begin{equation}
  \alpha_1^{[1]}=A\,\frac{\phi_{\xi}}
  {\phi} +\alpha_1^{[0]}\ ,
  \quad\quad
  \alpha_2^{[1]}=B\,\frac{\phi_{\xi}}{\phi}+\alpha_2^{[0]}\ .
\label{eq:10}
\end{equation}
Equation~(\ref{eq:10}) is an auto-B\"{a}cklund transformation, where $\vc\alpha^{[0]}$ is the seed solution and $\vc\alpha^{[1]}$ the iterated one. Substitution of Eq.~(\ref{eq:10}) into Eq.~(\ref{eq:5}) yields polynomials in powers of $\phi$ whose coefficients should be zero. The cumbersome calculations can be handled with the aid of MAPLE. The result of this procedure are summarized in what follows.

\subsection*{Expressions of the fields in terms of the singular manifold}

An easier way to deal with equations is obtained if we define the following quantities
\begin{equation}
r=\frac{\phi_t}{\phi_{\xi}}\ ,\qquad
v=\frac{\phi_{{\xi}{\xi}}}{\phi_{\xi}}\ , \qquad
s=v_{\xi}-\frac{v^2}{2}\ ,
\label{eq:11}
\end{equation}
such that the expressions of the seed fields are
\begin{eqnarray}
\alpha_1^{[0]}&=&-A_{\xi}+i\pi\left(A-\gamma B\right)-\frac{A}{2}\left(v+ir\right)\ ,\nonumber\\
\alpha_2^{[0]}&=&-B_{\xi}-i\pi\left(B+\gamma A\right)-\frac{A}{2}\left(v+ir\right)\ .
\label{eq:12}
\end{eqnarray}

\subsection*{SMM equations}

The equations that the singular manifold should satisfy in order to fulfill the truncation can be written as
\begin{subequations}
\begin{align}
r&=-2\lambda+i\left(A^{\dag}A_{\xi}+B^{\dag}B_{\xi}\right)+\pi\left(AA^{\dag}-BB^{\dag}\right)
  \nonumber\\
  &-\gamma\pi\left(AB^{\dag}+A^{\dag}B\right)\ ,
\label{eq:13}
\end{align}
\begin{align}
&A_t=iA_{{\xi}{\xi}}+2\pi A_{\xi} -2\pi\gamma B_{\xi}-2i\pi^2\left(1+\gamma^2\right)A\nonumber\\
&+A\left[-r_{\xi}+iv_{\xi}-\frac{i}{2}r^2-\frac{i}{2}v^2-2i\left(A_{\xi}A^{\dag}_{\xi}+B_{\xi}B^{\dag}_{\xi}\right)\right]
\nonumber\\
&+2Ar\left(AA^{\dag}_{\xi}
+BB^{\dag}_{\xi}\right)-2iA\pi\gamma r\left(AB^{\dag}+BA^{\dag}\right)
\nonumber\\
&+2A\pi\gamma\left[AB^{\dag}_{\xi}-B^{\dag}A_{\xi}+BA^{\dag}_{\xi}-A^{\dag}B_{\xi}\right]
\nonumber\\
&-2A\pi\left[ ir\left(BB^{\dag}-AA^{\dag}\right)+2 \left(B^{\dag}B_{\xi}+AA^{\dag}_{\xi}\right)\right]\ ,
\label{eq:14}
\end{align}
and
\begin{align}
&B_t=iB_{{\xi}{\xi}}-2\pi B_{\xi} -2\pi\gamma A_x-2i\pi^2\left(1+\gamma^2\right)B\nonumber\\
&+B\left[-r_{\xi}+iv_{\xi}-\frac{i}{2}r^2-\frac{i}{2}v^2-2i\left(A_{\xi}A^{\dag}_{\xi}+B_{\xi}B^{\dag}_{\xi}\right)\right]\nonumber\\
&+2Br\left(AA^{\dag}_{\xi}+BB^{\dag}_{\xi}\right)-2iB\pi\gamma r\left(AB^{\dag}+BA^{\dag}\right)\nonumber\\
&+2B\pi\gamma\left[AB^{\dag}_{\xi}-B^{\dag}A_{\xi}+BA^{\dag}_{\xi}-A^{\dag}B_{\xi}\right]\nonumber\\
&+2B\pi\left[ -ir\left(BB^{\dag}-AA^{\dag}\right)+2 \left(A^{\dag}A_{\xi}+BB^{\dag}_{\xi}\right)\right]\ .
\label{eq:15}
\end{align}
\label{eq:12_14}
\end{subequations}

\subsection*{Eigenfunctions}

The set of Eqs.~(\ref{eq:12_14}) can be simplified by introducing three function $\psi(\xi,t)$, $\omega(\xi,t)$
and $\eta(\xi,t)$ through the following definitions
\begin{equation}
A=\frac{\omega}{\psi}\ ,
\qquad
B=\frac{\eta}{\psi}\ ,
\label{eq:16}
\end{equation}
that allow us to write the condition $AA^{\dag}+BB^{\dag}=1$ as follows
\begin{equation}
\omega\omega^{\dag}+\eta\eta^{\dag}-\psi\psi^{\dag}=0\ .
\label{eq:17}
\end{equation}
From Eqs.~(\ref{eq:12_14}) we have
\begin{eqnarray}
&&v=\frac{\psi_{\xi}}{\psi}+\frac{\psi^{\dag}_{\xi}}{\psi^{\dag}},\nonumber\\
&&r=-2\lambda-i\left(\frac{\psi_{\xi}}{\psi}-\frac{\psi^{\dag}_{\xi}}{\psi^{\dag}}\right)\ .
\label{eq:18}
\end{eqnarray}
According to Eq.~(\ref{eq:11}), the singular manifold can be obtained by integration of the differential
\begin{equation}
d\phi=\psi\psi^{\dag}d\xi-\left[2\lambda\psi\psi^{\dag}+i\left(\psi_{\xi}\psi^{\dag}
-\psi\psi^{\dag}_{\xi}\right)\right]dt\ .
\label{eq:19}
\end{equation}

\subsection*{Spatial part of the Lax pair}

If we define the vector
\begin{equation}
\vc\Psi=\begin{pmatrix}\psi\\ \omega\\	\eta\end{pmatrix},\label{eq:20}
\end{equation}
the expressions Eq.~(\ref{eq:12}) combined with Eq.~(\ref{eq:17}) yield
\begin{equation}
\vc\Psi_{\xi}=V_1(\alpha_1,\alpha_2)\vc\Psi+i\lambda V_2\vc\Psi+i\pi V_3(\gamma)\vc\Psi\ ,
\label{eq:21}
\end{equation}
where
\begin{eqnarray}
&&V_1=
\begin{pmatrix}
        0& -\alpha_1^{\dag}&-\alpha_2^{\dag}\\
   -\alpha_1 & 0&0\\	
	   -\alpha_2 &0&0
        \end{pmatrix},\nonumber\\ &&V_2=
\begin{pmatrix}
       -1 &0&0\\
  0 &1&0\\	
	   0&0&1
        \end{pmatrix},\quad V_3=
\begin{pmatrix}
       0&0&0\\
  0 &1&-\gamma\\	
	   0&-\gamma&-1
        \end{pmatrix}.
\label{eq:22}
\end{eqnarray}

\subsection*{Temporal part of the Lax pair}

A similar result can be obtained for the time derivative of $\vc\Psi$
\begin{eqnarray}
\vc\Psi_t & = & iU_1(\alpha_1,\alpha_2)\vc\Psi+\pi U_2(\pi,\gamma,\alpha_1,\alpha_2)\vc\Psi\nonumber\\
&+&2\lambda U_3(\lambda,\alpha_1,\alpha_2)\vc\Psi\ ,
\label{eq:23}
\end{eqnarray}
where
\begin{equation}
U_1(\alpha_1,\alpha_2)=
\begin{pmatrix}
\alpha_1\alpha_1^{\dag}+\alpha_2\alpha_2^{\dag}&
	(\alpha_1^{\dag})_{\xi}&
				(\alpha_2^{\dag})_{\xi}\\
-(\alpha_1)_{\xi}&-\alpha_1\alpha_1^{\dag}&-\alpha_1\alpha_2^{\dag}\\	
-(\alpha_2)_{\xi} &-\alpha_2\alpha_1^{\dag}&
-\alpha_2\alpha_2^{\dag}
        \end{pmatrix},\nonumber
				\end{equation}
                \begin{equation}		
                U_2(\pi,\gamma,\alpha_1,\alpha_2)=\begin{pmatrix}
   - i\pi(1+\gamma^2)&
\gamma\alpha_2^{\dag}-\alpha_1^{\dag}&	\gamma\alpha_1^{\dag}+\alpha_2^{\dag}\\
\gamma\alpha_2 -\alpha_1&0&0\\	
\gamma\alpha_1 +\alpha_2&0&0
        \end{pmatrix},\nonumber
\end{equation} 
\begin{equation}
			U_3(\lambda,\alpha_1,\alpha_2)=\begin{pmatrix}
 i\lambda&
\alpha_1^{\dag}&\alpha_2^{\dag}\\
\alpha_1&-i\lambda&0\\	
\alpha_2&0&-i\lambda
        \end{pmatrix}.
\label{eq24}
\end{equation}
Eq.~(\ref{eq:21}) and Eq.~(\ref{eq:23}) constitute a three component Lax pair \cite{latchio,ablowitz}, whose compatibility condition yields Eq.~(\ref{eq:5}).

\subsection*{Darboux transformations}

One of the main advantages of the above described SMM is that it allows us to construct an iterative procedure to obtain highly non-trivial solutions by means of the eigenfunctions of a trivial seed solutions. This method have been described and successfully applied in Refs.~\cite{Estevez07,Estevez16}. Let
\begin{equation}
\vc\alpha^{[0]}=\begin{pmatrix}\alpha_1^{[0]}\\ 
\alpha_2^{[0]}\end{pmatrix}\nonumber,
\end{equation} 
be a seed solution of Eq.~(\ref{eq:5}) and
\begin{equation}
\vc\Psi_j^{[0]}=\begin{pmatrix}\psi_j^{[0]}\\ \omega_j^{[0]}\\	\eta_j^{[0]}\end{pmatrix}\ , 
\qquad
j=1,2\ ,
\label{eq:25}
\end{equation}
two eigenvectors of the Lax pair associated to $\vc \alpha^{[0]}$ with eigenvalues $\lambda_j$. These Lax pairs are
\begin{eqnarray}
&&(\vc\Psi_j^{[0]})_{\xi}=V_1\left(\vc\alpha^{[0]}\right)\vc\Psi_j^{[0]}+i\lambda_j V_2\vc\Psi_j^{[0]}+i\pi V_3(\gamma)\vc\Psi_j^{[0]}\nonumber\\
&&(\vc\Psi_j^{[0]})_t=iU_1\left(\vc\alpha^{[0]}\right)\vc\Psi_j^{[0]}+\pi U_2\left(\pi,\gamma,\vc\alpha^{[0]}\right)\vc\Psi_j^{[0]}\nonumber\\&&\quad\quad\quad+2
\lambda_j U_3\left(\lambda_j,\vc\alpha^{[0]}\right)\vc\Psi_j^{[0]}\ ,
\qquad j=1,2\ .
\label{eq:26}
\end{eqnarray}
The associated singular manifolds are defined through the following exact derivatives [see Eq.~(\ref{eq:19})]
\begin{eqnarray}
d\phi_j^{[0]}&&=\left(\psi_j^{[0]}\right)\left(\psi_j^{[0]}\right)^{\dag}(d\xi-2\lambda_jdt)\nonumber \\ 
&&-i\left[\left(\psi_j^{[0]}\right)_x{\left(\psi_j^{[0]}\right)^{\dag}}-{\left(\psi_j^{[0]}\right)^{\dag}_x}
\left(\psi_j^{[0]}\right)\right]dt\ .
\label{eq:27}
\end{eqnarray}
According to Eq.~(\ref{eq:10}), we can use the singular manifold $\phi_1^{[0]}$ to construct an iterated solution $\vc \alpha^{[1]}$ in the following form
\begin{eqnarray}
&&\alpha_1^{[1]}=\alpha_1^{[0]}+\frac{\omega_1^{[0]}\left(\psi_1^{[0]}\right)^{\dag}}{\phi_1^{[0]}}\ ,\nonumber\\ 
&&\alpha_2^{[1]}=\alpha_2^{[0]}+\frac{\eta_1^{[0]}\left(\psi_1^{[0]}\right)^{\dag}}{\phi_1^{[0]}}\ .
\label{eq:28}
\end{eqnarray}

Equation~(\ref{eq:10}) provides the following expressions for the modulus 
\begin{eqnarray}
&&\left(\alpha_1^{[1]}\right)\left(\alpha_1^{[1]}\right)^{\dag}=\left(\alpha_1^{[0]}\right)\left(\alpha_1^{[0]}\right)^{\dag}\nonumber \\ &&\quad\quad+\frac{i\pi\gamma\left[\left(\omega_1^{[0]}\right)
\left(\eta_1^{[0]}\right)^{\dag}-
\left(\omega_1^{[0]}\right)^{\dag}
\left(\eta_1^{[0]}\right)\right]}{\phi_1^{[0]}}-\nonumber \\&&\quad\quad\left(\frac{\omega_1^{[0]}\left(\omega_1^{[0]}\right)^{\dag}}{\phi_1^{[0]}}\right)_{\xi}, \nonumber\\
&&\left(\alpha_2^{[1]}\right)\left(\alpha_2^{[1]}\right)^{\dag}=\left(\alpha_2^{[0]}\right)\left(\alpha_2^{[0]}\right)^{\dag}\nonumber\\&&\quad\quad-\frac{i\pi\gamma\left[\left(\omega_1^{[0]}\right)
\left(\eta_1^{[0]}\right)^{\dag}-
\left(\omega_1^{[0]}\right)^{\dag}
\left(\eta_1^{[0]}\right)\right]}{\phi_1^{[0]}}\nonumber\\ &&\quad\quad-\left(\frac{\eta_1^{[0]}\left(\eta_1^{[0]}\right)^{\dag}}{\phi_1^{[0]}}\right)_{\xi}\ .
\label{eq:29}
\end{eqnarray}
This iterated solution $\vc \alpha^{[1]}$ has also a Lax pair. Let
\begin{equation}
\vc\Psi_{1,2}^{[0]}={\displaystyle \begin{pmatrix}\psi_{1,2}^{[1]}\\ \omega_{1,2}^{[1]}\\		\eta_{1,2}^{[1]},\end{pmatrix}}
\label{eq:30}
\end{equation}
be an eigenvector for $\vc \alpha^{[1]}$ with spectral parameter $\lambda_2$ such that
\begin{eqnarray}
&&\left(\vc\Psi_{1,2}^{[1]}\right)_{\xi}=V_1\left(\vc\alpha^{[1]}\right)\vc\Psi_{1,2}^{[1]}+i\lambda_2 V_2\vc\Psi_{1,2}^{[1]}+i\pi V_3(\gamma)\vc\Psi_{1,2}^{[1]},\nonumber\\
&&\left(\vc\Psi_{1,2}^{[1]}\right)_t=iU_1\left(\vc\alpha^{[1]}\right)\vc\Psi_{1,2}^{[1]}+\pi U_2\left(\pi,\gamma,\vc\alpha^{[1]}\right)\vc\Psi_{1,2}^{[1]}\nonumber\\&&\quad\quad\quad\quad+2
\lambda_2 U_3\left(\lambda_2,\vc\alpha^{[1]}\right)\vc\Psi_{1,2}^{[1]}\ ,\nonumber\\\label{eq:31}
\end{eqnarray}
which allows us to construct the following singular manifold $\phi_{1,2}$ by integrating
\begin{eqnarray}
d\phi_{1,2}^{[1]}&&=\left(\psi_{1,2}^{[1]}\right)\left(\psi_{1,2}^{[1]}\right)^{\dag}(d\xi-2\lambda_2dt)\nonumber\\&&-
i\left[\left(\psi_{1,2}^{[1]}\right)_x{\left(\psi_{1,2}^{[1]}\right)^{\dag}}-{\left(\psi_{1,2}^{[1]}\right)^{\dag}_x}\left(\psi_{1,2}^{[1]}\right)\right]dt\ .
\label{eq:32}
\end{eqnarray}

The Lax pair~(\ref{eq:31}) can be understood as a non-linear system that couples the field $\vc \alpha^{[1]}$ and the eigenvector $\vc\Psi_{1,2}^{[1]}$. This implies that the Painlev\'{e} expansion~(\ref{eq:28}) for the fields should be accompanied by a similar expansion for the eigenfunctions that can be written in the following form
\begin{eqnarray}
&&\psi_{1,2}^{[1]}=\psi_{2}^{[0]}-\psi_{1}^{[0]}\frac{\Delta_{1,2}}{\phi_1^{[0]}}\ ,\nonumber\\
&&\omega_{1,2}^{[1]}=\omega_{2}^{[0]}-\omega_{1}^{[0]}\frac{\Delta_{1,2}}{\phi_1^{[0]}}\ ,\nonumber\\
&&\eta_{1,2}^{[1]}=\eta_{2}^{[0]}-\eta_{1}^{[0]}\frac{\Delta_{1,2}}{\phi_1^{[0]}}\ .
\label{eq:33}
\end{eqnarray}
Substitution of~(\ref{eq:28}) and~(\ref{eq:33}) in~(\ref{eq:26}) yields
\begin{align}
&\Delta_{i,j}=\Delta\left(\vc\Psi_{i}^{[0]},\vc\Psi_{j}^{[0]}\right)\nonumber\\
&=i\frac{\left(\omega_{i}^{[0]}\right)^{\dag}\left(\omega_{j}^{[0]}\right)
+\left(\eta_{i}^{[0]}\right)^{\dag}\left(\eta_{j}^{[0]}\right)-\left(\psi_{i}^{[0]}\right)^{\dag}\left(\psi_{j}^{[0]}\right)}{2(\lambda_i-\lambda_j)}\ ,
\label{eq:34}
\end{align}
where $\vc\Psi_{j}^{[0]}$ is the eigenvector for the seed solution  $\vc \alpha^{[0]}$ with eigenvalue $\lambda_j$ as defined in (\ref{eq:26}). It is easy to see that a similar expansion could be applied to $\phi_{1,2}^{[1]}$ in  (\ref{eq:32}). The result is
\begin{equation}
\phi_{1,2}^{[1]}=\phi_{2}^{[0]}-\frac{\Delta_{1,2}\Delta_{1,2}^{\dag}}{\phi_{1}^{[0]}}.\label{eq:35}
\end{equation}

\subsection*{$\tau$--functions}

As far as $\phi_{1,2}^{[1]}$ is a singular manifold for $\vc \alpha^{[1]}$, we can iterate~(\ref{eq:28}) as
\begin{eqnarray}
&&\alpha_1^{[2]}=\alpha_1^{[1]}+\frac{\omega_{1,2}^{[1]}\left(\psi_{1,2}^{[1]}\right)^{\dag}}{\phi_{1,2}^{[1]}}\ ,
\nonumber\\
&&\alpha_2^{[2]}=\alpha_2^{[1]}+\frac{\eta_{1,2}^{[1]}\left(\psi_{1,2}^{[0]}\right)^{\dag}}{\phi_{1,2}^{[1]}}\ ,
\label{eq:36}
\end{eqnarray}
which combined with Eq.~(\ref{eq:23}) yields
\begin{eqnarray}
&&\alpha_1^{[2]}=\alpha_1^{[0]}+\frac{\omega_1^{[0]}\left(\psi_1^{[0]}\right)^{\dag}}{\phi_1^{[0]}}+\frac{\omega_{1,2}^{[1]}\left(\psi_{1,2}^{[1]}\right)^{\dag}}{\phi_{1,2}^{[1]}}\ ,\nonumber\\
&&\alpha_2^{[2]}=\alpha_2^{[0]}+\frac{\eta_{1}^{[0]}\left(\psi_1^{[0]}\right)^{\dag}}{\phi_1^{[0]}}+\frac{\eta_{1,2}^{[1]}\left(\psi_{1,2}^{[1]}\right)^{\dag}}{\phi_{1,2}^{[1]}}\ .
\label{eq:37}
\end{eqnarray}
Through the combination of Eq.~(\ref{eq:28}) with  Eq.~(\ref{eq:36}), we obtain the expressions of the second iteration in terms of the eigenfunctions of the seed equations
\begin{eqnarray}
&&\alpha_1^{[2]}=\alpha_1^{[0]}\nonumber \\ &&\quad\quad-\frac{\Delta_{1,2}^{\dag}\left(\psi_{1}^{[0]}\right)^{\dag}\left(\omega_{2}^{[0]}\right)+\Delta_{1,2}\left(\psi_{2}^{[0]}\right)^{\dag}\left(\omega_{1}^{[0]}\right)
}{\tau_{1,2}}\nonumber\\
&&\quad\quad+\frac{\phi_1^{[0]}\left(\psi_{2}^{[0]}\right)^{\dag}\left(\omega_{2}^{[0]}\right)+\phi_2^{[0]}\left(\psi_{1}^{[0]}\right)^{\dag}\left(\omega_{1}^{[0]}\right)
}{\tau_{1,2}}\ ,\nonumber\\
&&\alpha_2^{[2]}=\alpha_2^{[0]}\nonumber\\  && \quad\quad-\frac{\Delta_{1,2}^{\dag}\left(\psi_{1}^{[0]}\right)^{\dag}\left(\eta_{2}^{[0]}\right)+\Delta_{1,2}\left(\psi_{2}^{[0]}\right)^{\dag}\left(\eta_{1}^{[0]}\right)
}{\tau_{1,2}}\nonumber\\&&\quad \quad+ \frac{\phi_1^{[0]}\left(\psi_{2}^{[0]}\right)^{\dag}\left(\eta_{2}^{[0]}\right)+\phi_2^{[0]}\left(\psi_{1}^{[0]}\right)^{\dag}\left(\eta_{1}^{[0]}\right)
}{\tau_{1,2}}\ .\label{eq:38}
\end{eqnarray}
where we have defined the $\tau$--function $\tau_{1,2}$ as
\begin{equation}\tau_{1,2}=\phi_1^{[0]}\phi_2^{[0]}-\Delta_{1,2}\Delta_{1,2}^{\dag}\ .
\label{39}
\end{equation}

In conclusion, $\vc \alpha^{[0]}$ and its eigenfunctions $\vc \Phi_j^{[0]}$ allows us to obtain directly the first iterated solution $\vc \alpha^{[1]}$ as well as the second $\vc \alpha^{[2]}$ through Eq.~(\ref{eq:28}) and Eq.~(\ref{eq:36}) and the matrix $\Delta_{i,j}$ as defined in Eq.~(\ref{eq:34}). The following section is devoted to the use of this procedure to build up soliton solutions.

\section{Soliton solutions}

We start with the following trivial seed solution
\begin{equation*}
\vc\alpha^{[0]}=j_0e^{-2ij_0^2t}
\begin{pmatrix}
      \beta_{1}\\
\beta_{2}
        \end{pmatrix}.
\end{equation*}
$\beta_{1}$ and $\beta_{2}$ are parametrized as
\begin{equation}
\begin{pmatrix}
     \beta_{1}\\
   \beta_{2}
        \end{pmatrix}=\frac{1}{2}\begin{pmatrix}
     (1+s)\cos\theta_0+(1-s)\sin\theta_0\\
     (1-s)\cos\theta_0-(1+s)\sin\theta_0
\end{pmatrix}\ ,
\label{40}       
\end{equation}
where $s=\pm 1$. To deal with the focusing and defocusing cases together, we can leave $j_0$ as a free parameter. Actually $j_0$ should be real in the defocusing case and purely imaginary $j_0=ih_0$, in the focusing one. Solutions to the Lax pair~(\ref{eq:26}) are
\begin{eqnarray}
&&\psi_j=e^{k_j\left(\xi+c_jt\right)}e^{\frac{im_0\pi(\xi-m_0\pi t)}{2}}e^{ij_0^2t}\ ,\nonumber\\
&&\omega_j=d_je^{k_j\left(\xi+c_jt\right)}e^{\frac{im_0\pi(\xi-m_0\pi t)}{2}}e^{-ij_0^2t}\ ,
\nonumber\\
&&\eta_j=h_je^{k_j\left(\xi+c_jt\right)}e^{\frac{im_0\pi(\xi-m_0\pi t)}{2}}e^{-ij_0^2t}\ ,
\label{eq:41}
\end{eqnarray}
where
\begin{equation} 
\gamma=\tan(2\theta_0),\quad m_0=\frac{s}{\cos(2\theta_0)}\ .
\label{42}
\end{equation}

Furthermore, the constants $c_j$, $\lambda_j$ and  $k_j$ satisfy the relations
\begin{eqnarray}
&&c_j=m_0\pi-2\lambda_j\ ,\nonumber\\
&& k_j^2+\left(\lambda_j^2+\frac{m_0\pi}{2}\right)^2=j_0^2\ ,
 \label{eq:43}\end{eqnarray}
that allows us to introduce an angle $\theta_j$ such that
\begin{eqnarray}
&& \lambda_j=-\frac{m_0\pi}{2}+j_0\cos\theta_j\ ,\nonumber\\
&&k_j=j_0\sin \theta_j\ .
\label{44}
\end{eqnarray}
The coefficients $d_j$ and $h_j$ are 
\begin{equation} 
d_j=-\,i\beta_{1}e^{-i\theta_j}\ ,
\quad h_j=-\,i\beta_{2}e^{-i\theta_j}\ .
\label{eq:45}
\end{equation}

Equations (\ref{eq:32}) and (\ref{eq:34}) can be now  easily used in order to get 
\begin{eqnarray}
\phi_1&=&\frac{1}{2j_0\sin\theta_1}\left(a_1+E_1^2\right)\ ,\nonumber\\
\phi_2&=&\frac{1}{2j_0\sin\theta_2}\left(a_2+E_2^2\right)\ ,\nonumber\\
\Delta_{1,2}&=&\frac{\sin(\theta_2-\theta_1)+i\left[\cos(\theta_2-\theta_1)-1\right]}{2j_0(\cos\theta_1-\cos\theta_2)}\,E_1E_2\ ,\nonumber\\
\label{eq:46}
\end{eqnarray}
where $a_1$ and $a_2$ are arbitrary constants of integration and
\begin{equation}E_j=e^{{\displaystyle \left\{j_0\sin\theta_j\left[\xi+2(m_0\pi-j_0\cos\theta_j)t\right]\right\}}}\ ,
\label{eq:47}
\end{equation}
with $j=1,2$. The $\tau$--function defined in Eq.~(\ref{39}) can be explicitly written as
\begin{equation}
\tau_{1,2}=\frac{1}{4k_1k_2}\left(a_1a_2+a_2E_1^2+a_1E_2^2+A_{1,2}E_1^2E_2^2\right)\ ,
\label{eq:48}
\end{equation}
where
\begin{equation}
A_{1,2}=1-\frac{2\sin\theta_1\sin\theta_2
\left[1-\cos(\theta_1-\theta_2)\right]}{(\cos\theta_1-\cos\theta_2)^2}\ .
\label{eq:49}
\end{equation}

The first iteration can be now be obtained through~(\ref{eq:29}) as
\begin{eqnarray}
&& |\alpha_1^{[1]}|^2=\beta_{1}^2j_0^2
\left(1-\frac{1}{j_0^2}\left[\frac{(\phi_1)_{\xi}}{\phi_1}\right]_{\xi}\right)\ ,\nonumber\\ &&|\alpha_2^{[1]}|^2=\beta_{2}^2j_0^2\left(1-
\frac{1}{j_0^2}\left[\frac{(\phi_1)_{\xi}}{\phi_1}\right]_{\xi}\right)\ ,
\label{eq:50}
\end{eqnarray}
and the second iteration is deduced from~(\ref{eq:38}) as
\begin{eqnarray}
|\alpha_1^{[2]}|^2&=&\beta_{1}^2j_0^2\left(1-\frac{1}{j_0^2}\left[\frac{(\tau_{1,2})_{\xi}}{\tau_{1,2}}\right]_{\xi}\right)\ ,\nonumber\\ 
|\alpha_2^{[2]}|^2&=&\beta_{2}^2j_0^2\left(1-\frac{1}{j_0^2}
\left[\frac{(\tau_{1,2})_{\xi}}{\tau_{1,2}}\right]_{\xi}\right)\ .
\label{eq:51}
\end{eqnarray}

\subsection{Defocusing case ($g>0$). Dark solitons}

\begin{figure}[ht]
\includegraphics[width=0.75\columnwidth]{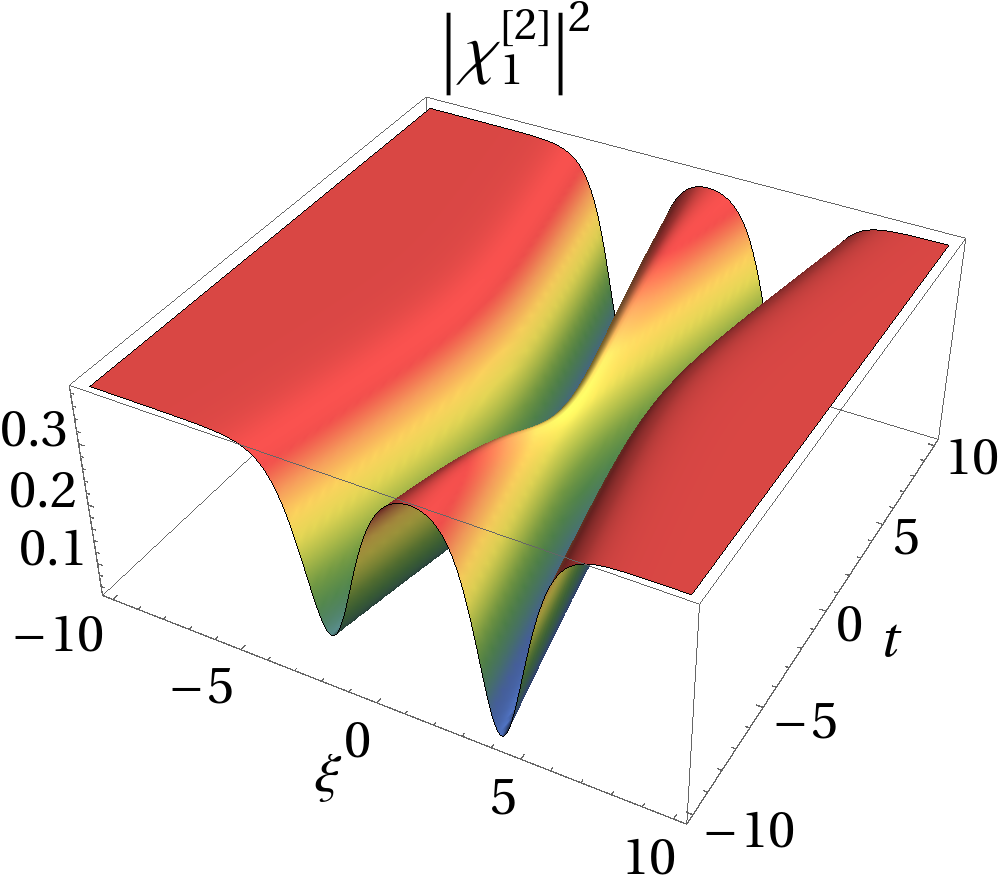} 
\caption{Squared modulus of the upper component of a two--soliton solution $|\chi_1^{[2]}|^2$ as a function of $\xi$ and $t$. Parameters are $g=2$, $\theta_0=0.5$, $\theta_1=1$, $\theta_2=1.2$, $s=1$, $j_0=1$, and $a_1=a_2=1$.}
\label{fig1}
\end{figure}

If we are dealing with $g>0$, $j_0$ should be a real parameter and, according to Eq~(\ref{eq:12}), we have dark solitons which, for the first iteration, are
\begin{eqnarray}
|\chi_1^{[1]}|^2&=&j_0^2\frac{1+s\cos(2\theta_0)}{2g}\left(1-\frac{1}{j_0^2}\left[\frac{(\phi_1)_{\xi}}{\phi_1}\right]_{\xi}\right)\ ,\nonumber\\ 
|\chi_2^{[1]}|^2&=&j_0^2\frac{1-s\cos(2\theta_0)}{2g}\left(1-\frac{1}{j_0^2}
\left[\frac{(\phi_1)_{\xi}}{\phi_1}\right]_{\xi}\right)\ ,
\label{52}
\end{eqnarray}
and the second iteration yields
\begin{eqnarray}
|\chi_1^{[2]}|^2&=&j_0^2\frac{1+s\cos(2\theta_0)}{2g}\left(1-\frac{1}{j_0^2}\left[\frac{(\tau_{1,2})_{\xi}}{\tau_{1,2}}\right]_{\xi}\right)\ ,\nonumber\\ 
|\chi_2^{[2]}|^2&=&j_0^2\frac{1-s\cos(2\theta_0)}{2g}\left(1-\frac{1}{j_0^2}
\left[\frac{(\tau_{1,2})_{\xi}}{\tau_{1,2}}\right]_{\xi}\right)\ .
\label{53}
\end{eqnarray}
Figure~\ref{fig1} displays the upper component $|\chi_1^{[2]}|^2$ of a two--soliton solution in the system of center of mass obtained through the Galileo transformation $\xi\rightarrow \xi-(c_1+c_2)t/2$. Notice that the lower component satisfies $|\chi_2^{[2]}|^2=|\chi_1^{[2]}|^2\big[1-s\cos(2\theta_0)\big]/\big[1+s\cos(2\theta_0)\big]$.

A particular case of  Eq.~(\ref{eq:48}) can be  obtained by setting $\theta_2=\pi-\theta_1$ and $a_1=a_2=\cos\theta_1$. In this case Eq.~(\ref{eq:48}) yields
\begin{eqnarray}
&&\tau_{1,2}\sim \cosh\left[2j_0^2\sin(2\theta_1)t\right]\nonumber
\\&&+\cos\theta_1\cosh\left[2j_0\sin\theta_1\left(\xi+\frac{2s\pi}{\cos(2\theta_0)}t\right)\right].
\label{eq:54}\end{eqnarray}

\subsection{Focusing case ($g<0$). Breathers}

As it has been  said before, for the focusing case, $j_0$ should be purely imaginary which means $j_0=ih_0$ with $h_0$ real. By using Eq.~(\ref{eq:13}), the second iteration yields bright solitons 
\begin{eqnarray}
|\chi_1^{[2]}|^2&=&h_0^2\frac{1+s\cos(2\theta_0)}{2|g|}\left(1+\frac{1}{h_0^2}\left[\frac{(\tau_{1,2})_{\xi}}{\tau_{1,2}}\right]_{\xi}\right),
\nonumber\\ 
|\chi_2^{[2]}|^2&=&h_0^2\frac{1-s\cos(2\theta_0)}{2|g|}\left(1+\frac{1}{h_0^2}
\left[\frac{(\tau_{1,2})_{\xi}}{\tau_{1,2}}\right]_{\xi}\right),
\label{eq:56}
\end{eqnarray}
where the equivalent of Eq.~(\ref{eq:48}) now is
\begin{eqnarray}
&&\tau_{1,2}\sim \cosh\left[2h_0^2\sin(2\theta_1)t\right]\nonumber\\
&&+\cos\theta_1\cos\left[2h_0\sin\theta_1\left(\xi+\frac{2s\pi}{\cos(2\theta_0)}t\right)\right],
\label{eq:57}
\end{eqnarray}
which is a solution periodic in $\xi$ and hyperbolic in $t$. This solution constitutes a generalization  of the Akhmediev's breather~\cite{Vishnu,Akhmediev2010}.

Furthermore, if we set $\theta_1=i\hat\theta_1$ in Eq.~(\ref{eq:57}), the result is
\begin{eqnarray}
&&\tau_{1,2}\sim \cos\left[2h_0^2\sinh(2\hat\theta_1)t\right]\nonumber\\
&&+\cosh\hat\theta_1\cosh\left[2h_0\sinh\hat\theta_1\left(\xi+\frac{2s\pi}{\cos(2\theta_0)}t\right)\right],
\label{eq:58}
\end{eqnarray}
which is a solution periodic in $t$ and hyperbolic in $\xi$. It is actually a generalization of the breather of Kutnesov-Ma~\cite{Akhmediev2012}.

\subsection{Focusing case ($g<0$). Rogue waves}

In the last years, rogue waves have been described as a curious type of waves that appears from nowhere and disappear without a trace. The well known Peregrine soliton~\cite{Akhmediev2010} is an example of rogue wave for the focusing NLS equation. In Ref.~\cite{Vishnu} rogue waves for the Manakov system have been obtained. In this section, we will derive this type of solutions to Eq.~(\ref{eq:5}).

\subsubsection{Case I}

It is easy to see, that there exists limiting cases of Eq.~(\ref{eq:49}) when $k_j=0$. These cases arises when $\theta _j$ is $0$ or $\pi$. The corresponding  eigenfunctions are

\begin{eqnarray}
&&\psi_1=e^{\frac{im_0\pi(\xi-m_0\pi t)}{2}}e^{ij_0^2t}\ ,
\nonumber\\
&&\omega_1=-i\beta_{1}e^{\frac{im_0\pi(\xi-m_0\pi t)}{2}}e^{-ij_0^2t}\ ,
\nonumber\\
&&\eta_1=-i\beta_{2}e^{\frac{im_0\pi(\xi-m_0\pi t)}{2}}e^{-ij_0^2t}\ .
\label{59}
\end{eqnarray}
when $\theta_1=0$, $ \lambda_1=-m_0\pi/2+j_0$ and $c_1=2\left(m_0\pi-j_0\right)$, and
\begin{eqnarray}
&&\psi_2=e^{\frac{im_0\pi(\xi-m_0\pi t)}{2}}e^{ij_0^2t}\ ,\nonumber\\
&&\omega_2=i\beta_{1}e^{\frac{im_0\pi(\xi-m_0\pi t)}{2}}e^{-ij_0^2t}\ ,
\nonumber\\
&&\eta_2=i\beta_{2}e^{\frac{im_0\pi(\xi-m_0\pi t)}{2}}e^{-ij_0^2t}\ .
\label{60}
\end{eqnarray}
for $\theta_2=\pi$, $ \lambda_2=-m_0\pi/2-j_0$ and $c_2=2\left(m_0\pi+j_0\right)$.

\begin{figure}[ht]
\includegraphics[width=0.75\columnwidth]{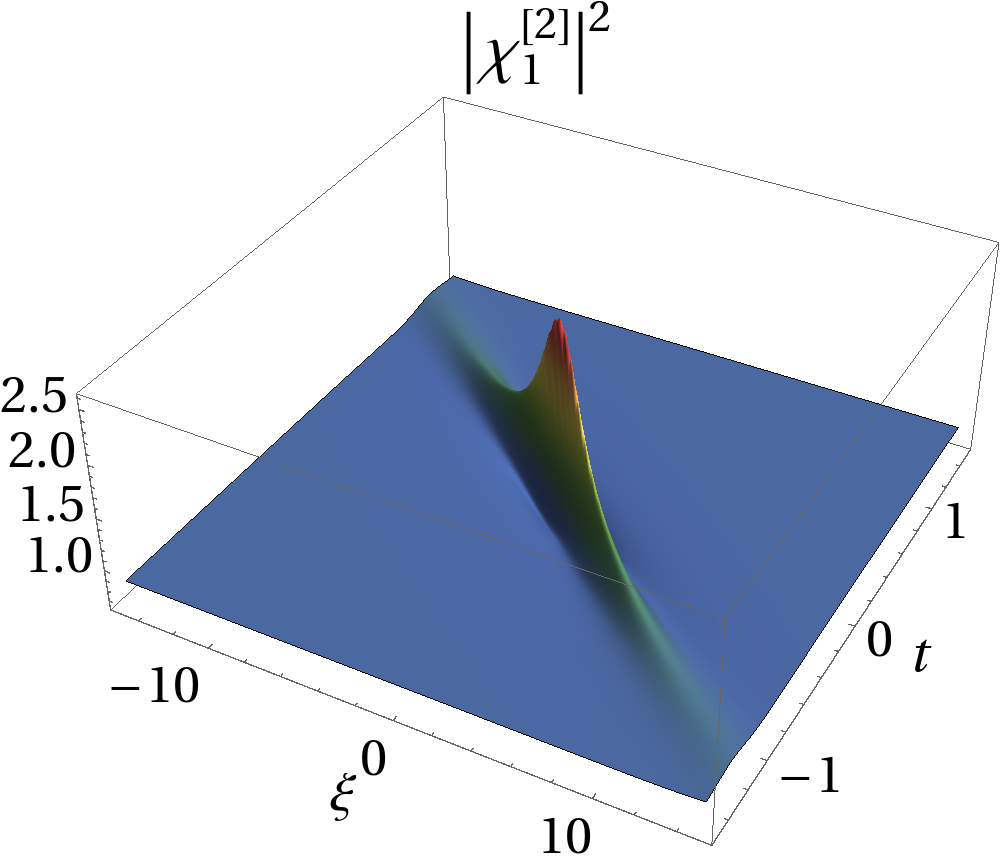}
\caption{Squared modulus of the upper component of a rogue wave I. Parameters are $g=-2$, $\theta_0=0.5$, $s=1$ and $h_0=1.5$.}
\label{fig2}
\end{figure}

For the focusing case, we take $j_0=ih_0$ ($h_0$ real) and obtain the following results
\begin{eqnarray}
&&\phi_1=\xi+2\pi m_0t-2ih_0t\ ,\nonumber\\
&&\phi_2=\xi+2\pi m_0t+2ih_0t\ ,\nonumber\\
&&\tau_{1,2}=(\xi+2\pi m_0t)^2+4h_0^2t^2+\frac{1}{h_0^2}\ .
\label{eq:61}
\end{eqnarray}
The behavior of the upper component for the above value of $\tau_{1,2}$ is presented in Fig.~\ref{fig2} for $g=-2$, $\theta_0=0.5$, $s=1$ and $h_0=1.5$. The lower component is obtained after rescaling the upper component and, therefore, it is not shown in the figure. This solution can be considered as a generalization of the Peregrine soliton~\cite{Akhmediev2010}.

\subsubsection{Case II}

It is straightforward to prove that there exists a slightly more complicated solutions to the Lax pair~(\ref{eq:25}). These solutions are
\begin{eqnarray}
&&\psi_1=\left(\xi+2\left(m_0\pi-j_0\right)t+\frac{i}{2j_0}\right)e^{\frac{im_0\pi(\xi-m_0\pi t)}{2}}e^{ij_0^2t},\nonumber\\
&&\omega_1=-i\beta_{1}\left(\xi+2\left(m_0\pi-j_0\right)t-\frac{i}{2j_0}\right)e^{\frac{im_0\pi(\xi-m_0\pi t)}{2}}e^{-ij_0^2t},
\nonumber\\
&&\eta_1=-i\beta_{2}\left(\xi+2\left(m_0\pi-j_0\right)t-\frac{i}{2j_0}\right)e^{\frac{im_0\pi(\xi-m_0\pi t)}{2}}e^{-ij_0^2t},\nonumber
\label{62}
\end{eqnarray}
when $\theta_1=0$, $ \lambda_1=-m_0\pi/2+j_0$ and $c_1=2\left(m_0\pi-j_0\right)$, and 
\begin{eqnarray}
&&\psi_2=\left(\xi+2\left(m_0\pi+j_0\right)t-\frac{i}{2j_0}\right)e^{\frac{im_0\pi(\xi-m_0\pi t)}{2}}e^{ij_0^2t},\nonumber\\
&&\omega_2=i\beta_{1}\left(\xi+2\left(m_0\pi+j_0\right)t+\frac{i}{2j_0}\right)e^{\frac{im_0\pi(\xi-m_0\pi t)}{2}}e^{-ij_0^2t},
\nonumber\\
&&\eta_2=i\beta_{2}\left(\xi+2\left(m_0\pi+j_0\right)t+\frac{i}{2j_0}\right)e^{\frac{im_0\pi(\xi-m_0\pi t)}{2}}e^{-ij_0^2t}\ ,
\nonumber
\label{63}
\end{eqnarray}
when $\theta_2=\pi$, $ \lambda_2=-m_0\pi/2-j_0$ and $c_2=2\left(m_0\pi+j_0\right)$.

\begin{figure}[ht]
\includegraphics[width=0.75\columnwidth]{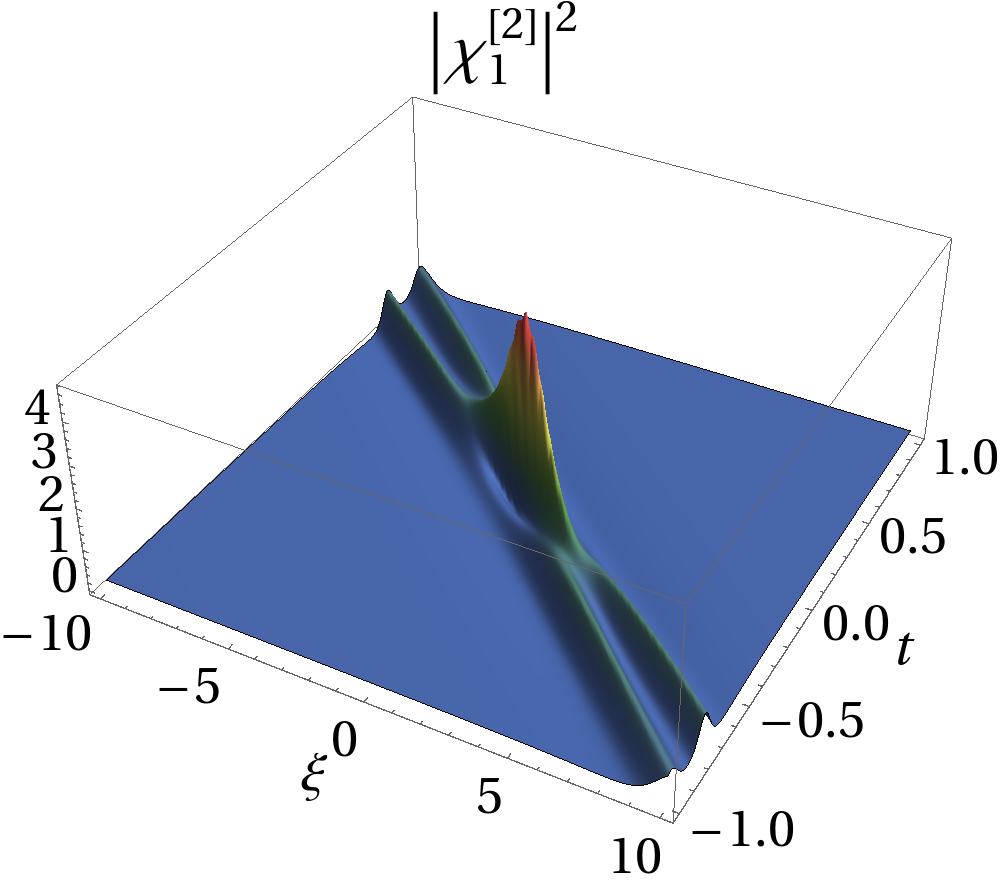}
\caption {Squared modulus of the upper component of a rogue wave II. Parameters are $s=1$, $g=-2$, $\theta_0=0.5$ and $h_0=0.7$.}
\label{fig3}
\end{figure}

For the focusing case, we should choose  $j_0=ih_0$ which yields the following results
\begin{eqnarray}
&&\phi_1=\left(\xi+2\pi m_0t\right)\left[\frac{1}{3}\left(\xi+2\pi m_0t\right)^2-4h_0^2t^2-\frac{1}{4h_0^2}\right]\nonumber\\
&&\quad +ih_0\left[-2t\left(\xi+2\pi m_0t\right)^2+\frac{3t}{2h_0^2}+\frac{8}{3}h_0^2t^3\right],\nonumber\\
&&\phi_2=\phi_1^{\dag},\nonumber\\
&&\tau_{1,2}=\left(\xi+2\pi m_0t\right)^2\left[\frac{1}{3}\left(\xi+2\pi m_0t\right)^2-4h_0^2t^2-\frac{1}{4h_0^2}\right]^2\nonumber\\&&\quad +h_0^2\left[-2t\left(\xi+2\pi m_0t\right)^2+\frac{3t}{2h_0^2}+\frac{8}{3}h_0^2t^3\right]^2\nonumber\\
&&\quad +h_0^2\left[\left(\xi+2\pi m_0t\right)^2+4h_0^2t^2+\frac{1}{4h_0^2}\right]^2.
\label{eq:64}
\end{eqnarray}
Figure~\ref{fig3} displays the upper component of the solution corresponding to Eq.~(\ref{eq:64}) for $s=1$, $g=-2$, $\theta_0=0.5$ and $h_0=0.7$. The lower component is obtained after rescaling the upper component and, therefore, it is not shown in the figure.

\subsection{Focusing case ($g<0$). Bright solitons}

Elliptic solutions to Eq.~(\ref{eq:3}) can be obtained through the ansatz
\begin{equation}
\vc\alpha^{[0]}=e^{-i\varphi\xi,t)}F(z)\begin{pmatrix}
      \beta_{1}\\
    \beta_{2}
        \end{pmatrix},\label{eq:65}
\end{equation}
where $z=\xi+ct$. The result is 
\begin{equation}
\varphi(\xi,t)=\frac{c}{2}\left(\xi+\frac{c}{2}t\right)-k^2t-\pi m_0(\xi+\pi m_0t)\ ,
\label{eq:666}
\end{equation}
and $F(z)$ obeys the elliptic equation
\begin{equation}F_{zz}-2F^3-k^2F=0,\nonumber\end{equation}
whose solution is
\begin{equation}
F(z)=\frac{km}{\sqrt{1-2m^2}}\,{\bf cn}\left(\frac{kz}{\sqrt{2m^2-1}};m\right)\ ,
\label{eq:67}
\end{equation}
where ${\bf cn}$ is the Jacobi elliptic cosine. The elliptic index $m$ and $k$ are arbitrary constants.
The hyperbolic limit $m=1$ yields, for the focusing case~(\ref{eq:13}), the solution
\begin{equation}
 \vc{\chi}=\sqrt{\frac{ 1}{ \mid g\mid}}\,\frac{k\,e^{-i\varphi(\xi,t)}}{\cosh\left[k(\xi+ct)\right]}\begin{pmatrix}
         e^{-i\pi(\xi+\pi t)}\,\beta_{1}\\
          e^{i\pi(\xi-\pi t)}\,\beta_{2}
 \end{pmatrix} \label{eq:68}.
\end{equation}
This is a solution that can be normalized by imposing $\int_{\infty}^{\infty} \vc \chi^{\dag}\cdot \vc \chi\, dx=1$. Therefore, the normalization implies $k=\mid g\mid/2$. For this value of $k$, the normalized solution can be finally written as 
\begin{equation}
\vc{\chi}=\frac{ \sqrt{\mid g\mid}}{2}\,\frac{e^{i\Omega(\xi,t)}}{\cosh\big[\mid g\mid (\xi+ct)/2\big]}\,\begin{pmatrix}
   e^{-i\pi\xi}\,\beta_{1}\\
   e^{i\pi\xi}\,\beta_{2}
\end{pmatrix}, 
\label{eq:69}
\end{equation}
where
\begin{align}
\Omega(\xi,t)&=-\left[\frac{c}{2}\left(\xi+\frac{c}{2}\,t\right)-\frac{g^2}{4}\,t\right]\nonumber\\
&+\pi\left[\frac{s\xi}{\cos(2\theta_0)}+\pi\gamma^2t\right]\ .  
\label{eq:70} 
\end{align}
This solution is the generalization of the Davydov soliton~\cite{Davydov79} that appears also in Ref.~\cite{Kartashov17}. Notice that the magnitude of the SOC is relevant only in the expression of the phase $\Omega(\xi,t)$. The generalized Davydov soliton~(\ref{eq:69}) has been recently put forward to stress the impact of the local deformation of the molecule about the carrier on spin-transport experiments~\cite{Diaz17bis}. In particular, it was found that the generalized Davydov soliton~(\ref{eq:70}) presents a well-defined spin projection onto the molecule axis that it is preserved during its motion, in spite of the fact that the electron spin is not a constant of motion of the linear Hamiltonian ($g=0$). Most importantly, the change of the chirality of the molecule, say from right-handed to left-handed, reverses the sign of spin projection of the generalized Davydov soliton~\cite{Diaz17bis}.

\bigskip

\section{Conclusions}

In this paper we have proposed a generalization of the Manakov system that includes an additional term with a SOC constant $\gamma$. Our model appears in the description of the spin dynamics in molecules with peptide dipoles in helical arrangement~\cite{Diaz17} or BECs with helical SOC in the absence of Zeeman splitting~\cite{Kartashov17}. We have studied the integrability of this model by means of the Painlev\'{e} test. The result is that the model has the Painlev\'{e} property and, therefore, it can be considered as an integrable model, as opposed to different generalizations that have not such property (e.g. model introduced by Kartashov and Konotop with non-vanishing Zeeman splitting~\cite{Kartashov17}). 

A direct consequence of the Painlev\'{e} property is the possibility of considering the singular manifold method as a powerful tool to derive many of the properties usually related to a non-linear integrable system. Actually, we have successfully used the singular manifold method to derive a three component Lax pair as well as binary Darboux transformations. These Darboux transformations easily yields the definitions of $\tau$-functions and an iterative method for the constructions of solutions. For the defocusing case, we have obtained dark solitons that generalize the solitons of the Manakov system. For the focusing case, breathers and rogue waves have been explicitly described. In addition, in this case case, solutions in terms of cnoidal waves also exist. The hyperbolic limit yields a generalization of the Davydov soliton whose spin projection onto the molecule axis depends on the chirality of the molecule~\cite{Diaz17bis}.

\acknowledgments

The authors thank R.~Guti\'{e}rrez for helpful discussions. This research has been supported by MINECO (Grants MAT2013-46308 and MAT2016-75955) and Junta de Castilla y Le\'{o}n (Grant SA045U16).

\bibliography{references}

\end{document}